\documentclass[prd,aps,showpacs,tightenlines,floats,nofootinbib,preprintnumbers]{revtex4}

\usepackage{graphicx}
\usepackage{amsmath,amssymb,color}

\newcommand{\dd}{\partial}

\newcommand{\xx}{\mathbf{x}}
\newcommand{\yy}{\mathbf{y}}
\newcommand{\kk}{\mathbf{k}}

\newcommand{\tr}{\hat{r}}

\newcommand{\HH}{\mathcal{H}}

\newcommand{\Mpl}{M_{\rm pl}}

\begin{document}

\title{Type-I cosmic string network}

\author{
Takashi Hiramatsu$^1$, 
Yuuiti Sendouda$^2$,
Keitaro Takahashi$^3$, 
Daisuke Yamauchi$^{4,5}$
and Chul-Moon Yoo$^{1,6}$
}

\affiliation{
$^1$ Yukawa Institute for Theoretical Physics, Kyoto University, Kyoto 606-8502, Japan\\
$^2$ Graduate School of Science and Technology, Hirosaki University, Hirosaki, Aomori 036-8561, Japan\\
$^3$ Faculty of Science, Kumamoto University, 2-39-1, Kurokami, Kumamoto 860-8555, Japan\\
$^4$ Institute for Cosmic Ray Research, The University of Tokyo, Kashiwa 277-8582, Japan\\
$^5$ Research Center for the Early Universe, School of Science, University of Tokyo, Tokyo, 113-0033, Japan\\
$^6$ Gravity and Particle Cosmology Group, Division of Particle and Astrophysical Science, Graduate School of Science, Nagoya University, Furo-cho, Chikusa-ku, Nagoya 464-8602, Japan
}

\preprint{YITP-13-56, RESCEU-36/13}

\begin{abstract}
We study the network of Type-I cosmic strings using the field-theoretic
 numerical simulations in the Abelian-Higgs model. For Type-I strings,
 the gauge field plays an important role, and thus we find  that the
 correlation length of the  strings is strongly dependent upon the
 parameter $\beta$, the ratio between the masses of the scalar field
 and the gauge field, namely,
 $\beta=m_\varphi^2/m_A^2$. In particular, if we take the cosmic
 expansion into  account, the network becomes densest in the comoving
 box for a specific value of $\beta$ for $\beta<1$.
\end{abstract}

\maketitle

\section{Introduction}

Cosmic strings are one-dimensional topological defects formed after
phase transitions. They are considered to make up a weblike structure
in the Universe, so-called \textit{the cosmic-string network}. Cosmic
strings could be a probe for the early phases of the Universe long
before the cosmic microwave background (CMB) epoch. They have a potential
to reveal the physics during the phase transition of fields in the early
Universe, and also be a potential source of 
gravitational waves \cite{Damour:2000wa,Damour:2001bk,Damour:2004kw,Siemens:2006yp,Siemens:2006vk,Abbott:2009rr,Abbott:2009ws,Olmez:2010bi,Kuroyanagi:2012wm}
and an extra source of CMB anisotropy 
\cite{Bevis:2007gh,Takahashi:2008ui,Yamauchi:2010ms,Yamauchi:2010vy,Yamauchi:2011cu,Dvorkin:2011aj,Urrestilla:2011gr,Kuroyanagi:2012jf,Battye:2010xz,Ade:2013xla}.

The simplest classical field-theoretic model to describe the string
formation is the Abelian-Higgs (AH) model, where there are a complex
scalar field with the self-coupling constant $\lambda$ and a
$U(1)$ gauge field with the gauge coupling constant $e$ (see e.g. the
textbook \cite{Vilenkin}). The basic properties of cosmic strings in the
AH model can be classified by a single parameter, 
$\beta=m_\varphi^2/m_A^2$, where $m_\varphi$ and $m_A$ are the
masses of the scalar field and the gauge field, respectively, acquired
after the spontaneous breaking of $U(1)$.
The case with $\beta=1$ is called ``critical coupling'' or 
``Bogomol'nyi coupling'', and the cases with $\beta<1$ and $\beta>1$ are
called Type-I and Type-II (cosmic) strings, respectively. These names
stemmed from the classification of superconductors and are not to be
confused with those of superstring theories. 

Historically, numerical simulations on the dynamical formation of the
string network in the expanding Universe have been performed in the
formulation based on the Nambu-Goto action (e.g.\ see \cite{Albrecht:1984xv,
Albrecht:1989mk, Bennett:1989yp, Allen:1990tv, Vincent:1996rb,
Martins:2005es, Ringeval:2005kr, Olum:2006ix, Fraisse:2007nu, BlancoPillado:2011dq}). 
In the Nambu-Goto simulations, strings are treated as zero-width
strings, and the detailed interactions between strings playing an
important role at the reconnection cannot be incorporated. Hence the
interactions of two strings are usually introduced by hand so that the
strings reconnect stochastically. 

On the bounty of the rapid development of computers, it has been becoming
possible to directly simulate the formation, evolution and extinction of
strings in the basis of the field-theoretic models on the lattice.
A pioneer work of the field-theoretic simulations was done in
Ref.~\cite{Vincent:1997cx}. After this work, some groups have tried to
perform simulation of the AH strings. 
Most studies assumed $\beta=1$. With this assumption, consistency
with one of the semianalytic models, velocity-dependent one-scale model,
was studied in \cite{Moore:2001px}, and the impacts of strings on the CMB were studied
in \cite{Bevis:2006mj, Bevis:2007gh, Bevis:2010gj, Mukherjee:2010ve}. 

Focusing on string interactions, Ref.~\cite{Bettencourt:1994kf}
clarified that there are no interactions between parallel straight
strings for $\beta=1$ in the Minkowski spacetime, while there is the
repulsive force between Type-II strings and the attractive force between
Type-I strings. Due to the attractive feature of the parallel Type-I
strings, they can form a bound state which could affect the
characteristics of the resultant network. In
Refs.~\cite{Bettencourt:1996qe, Salmi:2007ah}, the authors found the
nonintercommutation process accompanying a temporal bound state in the
collisions of strings with low velocity and small collision angle. A
similar feature can be seen in the cosmic superstring network
\cite{Sarangi:2002yt,Jones:2003da} where there are two kinds of strings,
F-strings and D-strings, and they can form a bound state called FD-strings 
\cite{Copeland:2003bj,Dvali:2003zj}. To investigate such a strong
interaction between strings, Urrestilla and Vilenkin have performed
simulations of scalar fields with a $U(1)\times U(1)$ gauge field and
they observed a small fraction of bound states formed in the string
network \cite{Urrestilla:2007yw}. As another interesting feature of
Type-I strings, it was reported that extremely high velocity
collisions also result in nonintercommutation \cite{Achucarro:2006es}.
These interesting characteristics could affect the Type-I string
network. As for the Type-II strings, there are a large number of network
simulations \cite{Pen:1997ae,Durrer:1998rw,Yamaguchi:1998gx,Yamaguchi:1999dy,
Yamaguchi:1999yp,Yamaguchi:2002zv,Yamaguchi:2002sh}, which have been
mainly used for studies on the cosmologically generated axions. Note
that the targets of these simulations are global strings corresponding to
the extreme Type-II case, $\beta\to\infty$. 

In most field theories including the AH model, coupling constants 
are expected to be of the same order. Therefore many of the previous
works on cosmic strings have dealt with the critical coupling or weakly
Type-I/II strings. However, as a special case, it is also reported that
a kind of minimally supersymmetric standard model prefers $\beta \ll 1$
(extreme Type-I strings) \cite{Cui:2007js}. Hence we stress that it is
still an open question what the preferred value of $\beta$ in the
Universe is, and field-theoretic simulations of the Type-I string
network including their interesting characteristics are needed. 

In this paper, we perform simulations of the Abelian-Higgs model with
various choices of $\beta<1$ in the radiation-dominated Universe. In an
expanding background, the strings feel additional effective repulsive
force, dragging effect by the cosmic expansion, in between them. Hence,
it is expected that the properties of the resultant string network
depend on not only the strength of the intrinsic attractive force
between strings, but also the Hubble parameter at the string formation
epoch. To investigate the characteristics of the network, we solve the
field equations of the scalar field and the gauge field in a numerical way.

This paper is organized by the following sections. In
Sec.~\ref{sec:basic}, we give the field equations to be solved and set 
up the model used throughout this paper. In Sec.~\ref{sec:setup}, we
define some numerical parameters for the following numerical
simulations. In Sec.~\ref{sec:identification}, we discuss how to find
the string cores and define the estimator of the correlation length.
In Sec.~\ref{sec:result}, we show numerical results of the network
simulations of Type-I strings varying the parameter $\beta$.
Then, in Sec.~\ref{sec:conclusion}, we conclude this paper.
In addition, we check the numerical convergence of our results in
Appendix~\ref{appsec:check}, and related with this, the $\beta$
dependence of the width of strings for their static configurations is
shown in Appendix~\ref{appsec:vortex}. Finally, in
Appendix~\ref{appsec:smoothing}, we explain our procedure to estimate
the effective number of strings in the horizon-sized box.

We use the units such that $\hbar=c=k_{\mathrm B}=1$.

\section{Basic equations}
\label{sec:basic}


In the spatially homogeneous and isotropic space-time, the 
metric $g_{\mu\nu}$ is parameterized by the scale factor $ a $ as 
%
\begin{equation}
ds^2 = g_{\mu\nu}dx^\mu dx^\nu = a^2(\tau)\eta_{\mu\nu}dx^\mu dx^\nu = a^2(\tau)(-d\tau^2+\delta_{ij}dx^idx^j).
\end{equation}
%
The action of the AH model is
%
\begin{equation}
 S = -\int\!d^4x\,\sqrt{-g}\left(\frac{1}{4}F_{\mu\nu}F^{\mu\nu} +
  (D_\mu\varphi)^*(D^\mu\varphi) + V(\varphi)\right), \label{eq:action}
\end{equation}
%
where the symbol $*$ denotes the complex conjugate and
%
\begin{equation}
 D_\mu \equiv \partial_\mu - ieA_\mu, \quad 
 F_{\mu\nu} \equiv \partial_\mu A_\nu - \partial_\nu A_\mu.
\end{equation}
%
Here we introduced the complex scalar field $\varphi$, and the $U(1)$
gauge field, $A_\mu$. Varying this action, we obtain the field equations for 
$\varphi$ and $A_\mu$ in arbitrary gauge as
%
\begin{align}
&\frac{1}{\sqrt{-g}}D_\mu(\sqrt{-g}g^{\mu\nu}D_\nu\varphi) =
\frac{\partial V}{\partial \varphi^*}, \label{eq:eqfullphi} \\
&\frac{1}{\sqrt{-g}}\partial_\mu(\sqrt{-g} F^{\mu\nu})
     = -2eg^{\mu\nu}{\rm Im}(\varphi^*D_\mu\varphi). \label{eq:eqfullA}
\end{align}
%
Throughout this paper, we take a gauge condition $A_0=0$.
Then the evolution equations to be solved in the Friedmann Universe become
%
\begin{align}
&\varphi'' + 2\HH\varphi' - \delta^{ij}D_iD_j\varphi + a^2\frac{dV}{d\varphi^*} = 0
\label{eq:eqphi},\\
&A''_i - \delta^{jk}\partial_j\partial_kA_i + \delta^{jk}\partial_i\partial_jA_k = 2ea^2{\rm Im}(\varphi^* \partial_i\varphi)-2e^2a^2A_i|\varphi|^2, \label{eq:eqA}
\end{align}
%
where the prime ( $'$ ) denotes the derivative with respect to the
conformal time $\tau$, $\HH$ is the comoving Hubble parameter defined as $\HH=aH=a'/a$, and
%
\begin{equation}
\delta^{ij}D_iD_j\varphi = \delta^{ij}\partial_i\partial_j\varphi
   -2ie\delta^{ij}A_i\partial_j\varphi
   -ie\delta^{ij}\partial_iA_j\varphi
   -e^2|A|^2\varphi \label{eq:DD}.
\end{equation}
%
The constraint equation given
by the $\nu=0$ component of Eq.~(\ref{eq:eqfullA}) becomes
%
\begin{equation}
\delta^{ij}\dd_i{A_j}' = 2ea^2{\rm Im}(\varphi^*\varphi'). \label{eq:const}
\end{equation}
%
In our simulations, we impose the periodic boundary condition on the
boundaries of the computational domain. Therefore the volume integral of
Eq.~(\ref{eq:const}) is trivially satisfied, and hence we do not
consider this equation hereafter.

We consider the following temperature-dependent potential:
%
\begin{equation}
V(\varphi;T) = 
 \frac{\lambda}{2}(\varphi^*\varphi - \eta^2)^2 +
 \frac{\lambda}{3}T^2\varphi^*\varphi. 
\end{equation}
%
The transition temperature $T_c=\sqrt{3}\eta$ is found as the
temperature at which the effective mass of 
$\varphi$ deriving from the second derivative of the potential
vanishes. After the phase transition, the scalar field starts to
oscillate around the true vacuum given by
%
\begin{equation}
 |\varphi_{\rm vac}| = \eta\sqrt{1-\left(\frac{T}{T_c}\right)^2}.
\end{equation}
%

The masses of the scalar field $\varphi$ and gauge field
$A_i$ after the phase transition are given as $m_\varphi^2 =
2\lambda\eta^2$ and $m_A^2 = 2e^2\eta^2$, respectively, in
the zero temperature limit. The ratio of these masses, 
%
\begin{equation}
 \beta \equiv \frac{m_\varphi^2}{m_A^2} = \frac{\lambda}{e^2},
\end{equation}
%
plays an important role for the characteristics of strings and the string
network at the late time\footnote{Note that the 
definition of $\lambda$ in this paper is different from that in the
literature, where the mass of the scalar field is calculated as
$m_\varphi^2 = \lambda\eta^2$ and hence the parameter $\beta$ is written by 
$\beta=\lambda/2e^2$.}. 
In this paper, to investigate the string
network constituted by the Type-I strings, we set $\beta \leq 1$.

\section{Simulation setup}
\label{sec:setup}

Throughout this paper, we assume the radiation-dominant Universe and
no backreaction from the scalar and gauge fields onto the background
geometry. Then the Friedmann equation is simply given by
%
\begin{equation}
 H = \frac{a'}{a^2} = \frac{T^2}{2\gamma\Mpl},
\qquad 
 \gamma \equiv \left(\frac{45}{16\pi^3g_*}\right)^{1/2},
 \label{eq:Friedmann2}
\end{equation}
%
where $g_*$ is the effective massless degrees of freedom. 

The vacuum expectation value of the scalar field at the zero
temperature $\eta$ determines the normalization of the typical energy
scales of $H$, $T$ and also the time scale and the spatial scale. 
For convenience, we introduce another parameter $q$ 
defined as
$q=\eta/(\gamma\Mpl)$. We fix $q=0.1$ throughout this paper
except in the last part of Sec.~\ref{subsec:correlation}.
Note that the constant $\gamma$ does not appear anywhere except in
Eq.~(\ref{eq:Friedmann2}). Hence it is not needed to set a specific
value to $\gamma$ or $g_*$.

In what follows the symbols with the subscript $i$ ($f$) refer to
the quantities at the initial (final) time of each simulation. 

Normalizing the initial value as $a(\tau_i)=1$, the scale factor can be
written as $a(\tau)=\tau/\tau_i=T_i/T$. The Hubble parameter is
then recast as $H = \tau_i/\tau^2$. The comoving
Hubble is denoted as $\HH$ and is given by $\HH=aH=1/\tau$. 

The initial, final and transition times are, respectively, given by 
%
\begin{equation}
\tau_i = \frac{2\eta}{qT_i^2},
\quad
\tau_f = \tau_i\left(\frac{T_i}{T_f}\right),
\quad
\tau_c = \tau_f\left(\frac{T_f}{T_c}\right).
\end{equation}
%
Our numerical simulations are performed before the phase transition and
end sufficiently after it, namely, $T_i>T_c$ and $T_f<T_c$.

The computational domain is a comoving box with the side length $L$, and
we define a new quantity 
%
\begin{equation}
 s(\tau) \equiv \frac{aL}{H^{-1}} = \frac{L}{\tau}
\end{equation}
%
to measure the relative size of the box at a given time compared to
the horizon scale.  We use $s_i=s(\tau_i)$ and $s_f=s(\tau_f)$ as
numerical parameters instead of the box size $L$ and the final time
$\tau_f$. In particular, we choose $s_f=2$ to avoid the contamination
from the finiteness of the computational domain throughout the simulations.

In the end, we are left with the physical parameters
$\{\lambda,e,\beta\}$, two of them being independent. 

As for the initial conditions, we set the values of $\varphi$ on the
assumption that the scalar field stays in the thermal bath with the
temperature $T$
\cite{Yamaguchi:1999yp}. Reference~\cite{Yamaguchi:1999yp} provides the
equal-time correlation function of $\varphi(\xx,\tau)$.
Subtracting the contribution from the infinite vacuum energy, we obtain
the equal-time correlation function of $\varphi(\xx,\tau)$,
%
\begin{gather}
 \langle T|\varphi(\xx,\tau)\varphi(\yy,\tau)^*|T\rangle_{\rm w/o\,vacuum}
 = \int\frac{d^3\kk}{(2\pi)^3\omega_k}
   \frac{1}{e^{\omega_k/T}-1}e^{i\kk\cdot(\xx-\yy)}, 
   \label{eq:initpower} 
\end{gather}
%
where $\omega_k=\sqrt{\kk^2+m^2}$, and 
$\langle T|\cdots|T\rangle_{\rm w/o\,vacuum}$
represents the ensemble average at the finite temperature
without the contribution from the vacuum energy.
The integrand
in the right-hand side gives the power spectra of $|\varphi(\kk)|$.
If we take a limit, $|\xx-\yy|\to 0$ and $m\to 0$, we obtain the
variances, $\langle|\varphi|^2\rangle=T^2/12$. 
In our simulations, we use Eq.~(\ref{eq:initpower}) with $m=0$.

Firstly we give the
initial condition for $\varphi(\kk,\tau_i)$ at each grid
point, $\mathbf{k}$, in the
Fourier space by generating the Gaussian random numbers for
$|\varphi(\kk,\tau_i)|$ according to the above power spectrum, and the
homogeneous random numbers between $0$ and $2\pi$ for the phase
of $\varphi(\kk,\tau_i)$. 
Then, using the inverse Fourier transformation, we obtain the initial
condition for $\varphi(\xx,\tau_i)$ in the real space.
As for the time derivative, $\varphi'(\xx,\tau_i)$, we simply set
$\varphi'=0$ everywhere. Next, $A'_i$ is determined from the constraint
equation with the Fourier transformation given in Eq.~(\ref{eq:const}),
while $A_i$ is set to be zero. We expect that the details of the initial 
condition do not crucially affect the final behavior of strings after
the phase transition. 

We use the staggered grid where $\varphi(\xx,\tau)$ lies at the grid points;
$A_i(\xx,\tau)$ connects at each link two grid points, $\xx$ and
$\xx + h\hat{\bf e}_i$, where $h$ is the grid spacing; and $\hat{\bf e}_i$
represents the unit vector parallel to the axis of $i (=x,y,z)$.
The fiducial values of the numerical parameters are listed in
Table~\ref{tab:param}. We use these values in most of our simulations,
except when we check the reliability of our numerical results,
dependence on the resolution and the box size.
In solving Eqs.~(\ref{eq:eqphi})(\ref{eq:eqA}) for $\varphi$ and
$A_i$, we use the 
second-order finite difference scheme for spatial derivatives and the second-order
leapfrog scheme for time evolution. 
To implement this time-evolution scheme, we define $\psi=a\varphi$ and
eliminate the first-derivative term, $\varphi'$ or $\psi'$.
Then we solve the equation of $\psi$, instead of $\varphi$.

In addition, in order to compare with the previous studies
\cite{Vincent:1997cx, Moore:2001px, Bevis:2006mj, Bevis:2010gj}, 
we implement the Press-Ryden-Spergel (PRS) algorithm \cite{Press:1989yh}
in the last part of Sec.~\ref{sec:result}. In the PRS prescription, the
coupling constants, $\lambda$ and $e$, become time dependent, 
%
\begin{equation}
 \lambda(\tau) = \frac{\lambda_i}{a^2}, \quad e(\tau) = \frac{e_i}{a},
 \label{eq:PRS}
\end{equation}
%
where $\lambda_i$ and $e_i$ are the initial values, and note that
$\beta$ keeps its constancy during the simulations.
This algorithm is effective for the lattice simulations in the
expanding Universe since the expanding lattice spacing can forever, in
principle, follow the width of the strings.

\begin{table}[!ht]
\caption{The fiducial values of numerical parameters. The parameters
 enclosed in parentheses are determined from other ones.}
\label{tab:param}
\begin{tabular}{|c|c|c|}
\hline
Grid size           & $N$   & $1024^3$ \\
Initial box size    & $s_i$ & 18 \\
Final box size      & $s_f$ & 2 \\
(Comoving box size) & $L$   & 30$\eta^{-1}$ \\
Energy scale & $q$ & 0.1 \\
Initial temp.       & $T_i/T_c$ & 2 \\
(Final temp.)       & $T_f/T_c$ & 2/9 \\
\hline
\end{tabular}
\end{table}

\section{String identification}
\label{sec:identification}

\subsection{String cores}
\label{subsec:string_cores}

In order to measure the total length of the strings in the network, we
first have to identify the location of the cores of strings.
However there is great ambiguity in how we identify them. In
this paper, we implement the method developed in Ref.~\cite{Yamaguchi:2002sh}.

Here we briefly explain this algorithm. Let us consider a cube
constituted by eight neighbor grid points  on which the scalar field
stays. Focusing on one of six surfaces of the cube, if the phase
of the scalar 
field on the surface becomes monotonically larger or smaller along its
four sides and eventually the 
sum of the differences of phases between two neighbor grid
points on the surface comes to $2\pi$, we
can judge that a string passes the surface in principle.

Instead of this direct method, we consider
a complex plane representing the complex scalar field, and divide it
into three domains. According to Ref.~\cite{Yamaguchi:2002sh}, 
there is an advantage to dividing the plane inhomogeneously as
$0<{\rm Arg}(\varphi)<\pi/2$, $\pi/2<{\rm Arg}(\varphi)<\pi$, and
$\pi<{\rm Arg}(\varphi)<2\pi$.
On this plane, we plot four values of $\varphi/|\varphi|$ on a surface of
the unit grid cube. We judge that a string passes on the surface if each
domain possesses at least one $\varphi/|\varphi|$. We repeat this
process for six surfaces of the cube, and determine the surfaces
across which the string passes. Then we connect the central points of
the surfaces by straight line segments. (Therefore, the length of a line
segment is $h$ or $h/\sqrt{2}$ where $h$ is the grid spacing.)
After repeating this procedure
for all cubes, we can identify the location of strings in the simulation box.
Finally, the total length of the strings can be estimated by summing the
lengths of the line segments.

\subsection{Correlation length}

In Ref.~\cite{Hindmarsh:2008dw}, the authors used an estimator of the
comoving correlation length of the string network by measuring the total
length of the strings in the simulation box, 
%
\begin{equation}
\xi =\sqrt{\frac{\mathcal V}{ L_{\rm str}}},  \label{eq:def_xi}
\end{equation}
%
where $\mathcal V$ is the comoving volume of the simulation box, and
$L_{\rm str}$ is the total comoving length of strings in the box, estimated
with the method discussed in the previous section.
The quantity $\xi$ represents the averaged comoving
separation of two neighboring strings, and thus is identical to the
correlation length from the viewpoint of the so-called one-scale model.
If the scaling regime is reached, $\xi$ should grow proportionally to the
conformal time, $\xi \propto \HH^{-1} =\tau$.
In this paper, we estimate the correlation length from simulations using
this estimator. 

\subsection{String energy}

If the scaling regime starts, the string energy should behave as
%
\begin{equation}
 \rho_{\rm str} = \frac{\mu \xi_p}{\xi_p^3} \propto \frac{\HH^2}{a^2} \propto
  \tau^{-4},
\end{equation}
%
where $\xi_p \equiv a\xi$ is the physical correlation length.
In order to directly see whether the network really gets to the scaling 
regime, we define and calculate the string energy and check whether it evolves 
according to this scaling.

Given an action $S$, the energy density can be derived as
%
\begin{equation}
\rho = u^\mu u^\nu T_{\mu\nu},\quad
T_{\mu\nu} = \frac{-2}{\sqrt{-g}}\frac{\delta S}{\delta g^{\mu\nu}},
\end{equation}
%
where $u^\mu=(a^{-1},0,0,0)$ denotes the comoving observer's four-velocity.
Using Eq.~(\ref{eq:action}), this can be expressed as
%
\begin{equation}
 \rho = \frac{1}{a^2}\left[
    |\varphi'|^2 + |\partial_i\varphi|^2 + e^2|A_i|^2|\varphi|^2
    -2eA^i \mathrm{Im}(\varphi^*\partial_i\varphi) + a^2V
           \right]
       + \frac{1}{2a^4}\left(
         |A'_i|^2 + F_{12}^2+F_{23}^2+F_{31}^2
       \right). \label{eq:energy}
\end{equation}
%

Next we consider the width of a string to estimate the energies
possessed by each string. In this paper, we do not directly calculate
the width, but recognize the region for $|\varphi|<\varphi_c$ with a given constant $\varphi_c$,
e.g.\ $\varphi=0.5\eta$, as a portion of strings, namely,
%
\begin{equation}
  \rho_{\rm str} \equiv \frac{1}{\mathcal{V}}\int_{\mathcal{V}}
      \rho(\xx,\tau)\Theta(\varphi_c-|\varphi(\xx,\tau)| ) \,d^3\xx,
 \label{eq:rhostr}
\end{equation}
%
where $\Theta(x)$ is the Heaviside step function.
 
\section{Results}
\label{sec:result}

\subsection{Basic behavior of string network}

\begin{figure}[!ht]
\centering{
  \includegraphics[scale=0.55]{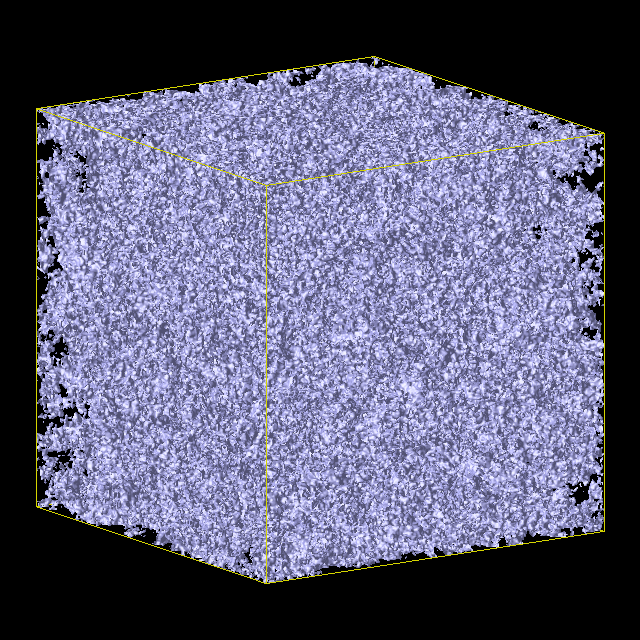}
  \includegraphics[scale=0.55]{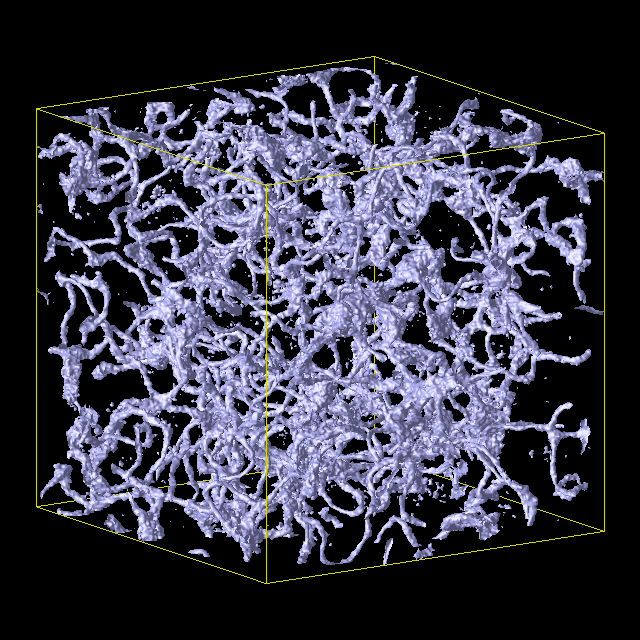}
  \includegraphics[scale=0.55]{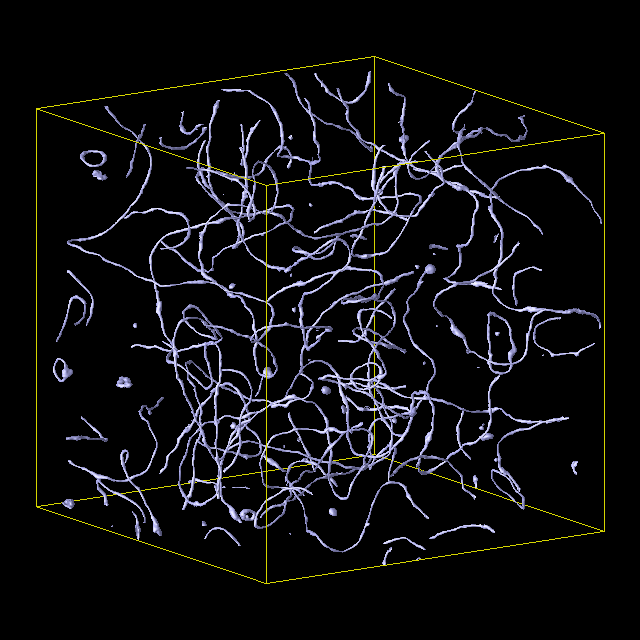}
}
\caption{The isosurfaces with $|\varphi|=0.5\eta$ at $\tau\eta=5.45,\;6.27$ and
 $11.53$ for the simulation with $\lambda=1.0$ and $\beta=0.2$.}
\label{fig:3D}
\end{figure}

In Fig.~\ref{fig:3D}, we plotted three different time slices of the
scalar field with $\lambda=1.0$ and $\beta=0.2$ as an example,
which shows 
the isosurface with $|\varphi|=0.5\eta$. From left,
they correspond to  $\tau\eta=5.45,\;6.27$, and $11.53$, respectively. 
Note that the phase transition takes place at
$\tau_c\eta=10/3$ in this simulation. 

After the phase transition, the scalar field starts to oscillate around
the true vacuum in the so-called Mexican hat potential. Just after the
transition, the oscillation is still so strong that it is not obvious
whether the strings are actually forming, as shown in the left panel of
Fig.~\ref{fig:3D}. After a while, due to the Hubble friction, the scalar
field in most parts of the computational box is settling down to
the true vacuum, and satisfies $V(\varphi;T)<V(0;T)$. Then the strings
begin to appear, as shown in the center panel of Fig.~\ref{fig:3D}. 
At this phase, we observed the pulsating behavior of strings: that is,
the width of strings oscillates in time. This feature can also be
seen in the time evolution of string energy shown in the next subsection. 

At the later time, strings clearly appear in the box, as shown in the right
panel of Fig.~\ref{fig:3D}. Even at this time, the string pulsation can
be still observed, and for some strings, we found a phenomenon such
that a wave packet is moving on a string, for instance, on a small loop
with a clump shown in the left center of the box.

\subsection{String energy}

Next we identify all strings at each time step, and calculate their
energies to check whether the string network is in the scaling regime.
To calculate their energies, we do not use the string core identification
technique discussed in Sec.~\ref{subsec:string_cores} since it is difficult to set
the string width which changes in time, and it does not matter where
the string core is. Instead, we fix $\varphi_c$, and
we identify the region satisfying $|\varphi|<\varphi_c$ as strings.
Then the string energy is given by Eq.~(\ref{eq:rhostr}).

\begin{figure}[!ht]
\centering{
 \includegraphics[width=10cm]{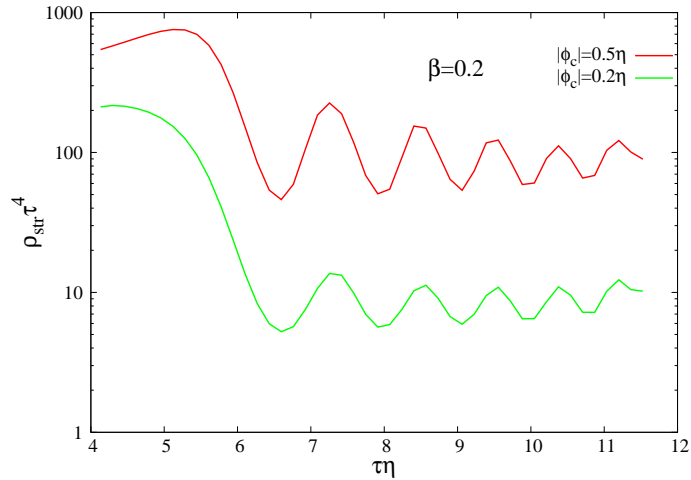}
}
\caption{Time evolution of the string energy for $\beta=0.2$ and
 $\lambda=1.0$. The red line and green line correspond to
 $\varphi_c=0.5\eta$ and $\varphi_c=0.2\eta$, respectively.} 
\label{fig:energy}
\end{figure}

In Fig.~\ref{fig:energy}, we show the time evolution of the
string energy for $\beta=0.2$ and $\lambda=1.0$ with $\varphi_c=0.5\eta$
and $\varphi_c=0.2\eta$. For $\tau\eta \gtrsim 6.5$, we find that the
relation $\rho_{\rm str}\propto \tau^{-4}$ is approximately satisfied at
the late time, if we average the oscillatory behavior, and that the
transition time seems insensitive to the choice of $\varphi_c$.
This observation indicates that the network would lie in the scaling
regime. 

\subsection{Correlation length}
\label{subsec:correlation}

\begin{figure}[!ht]
\centering{
 \includegraphics[width=10cm]{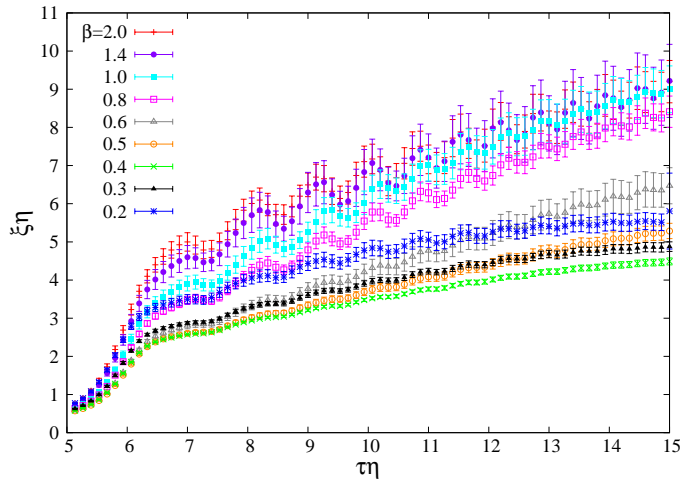}
}
\caption{Time evolution of the correlation length estimated by Eq.~(\ref{eq:def_xi}).}
\label{fig:corr}
\end{figure}

In Fig.~\ref{fig:corr}, we show the time evolution of the correlation
length estimated by Eq.~(\ref{eq:def_xi}) for various values of $\beta$
including weakly Type-II regime with $\lambda=1.0$. Each of them
is the average of 10 realizations, and 
the error bars indicate one $\sigma$. We found that the correlation
length is strongly dependent upon $\beta$, particularly for 
$\beta < 1.0$. 

Our findings are as follows. First, the time of the start of the
string formation is common to all the cases and is about
$\tau\eta\approx 6.5$. Moreover, there is a general tendency that the 
correlation length $\xi$ becomes smaller. For $0.8 \leq \beta \leq 1.4$, 
$\xi|_{\tau\eta\approx 6.5}$ becomes smaller, which indicates the
initial density of the string network would be larger. For 
$0.4 \leq \beta \leq 0.8$, $\xi|_{\tau\eta\approx 6.5}$ does not 
change so much, but the increasing rate of $\xi$, or $d\xi/d\tau$,
becomes smaller as $\beta$ approaches $0.4$. Finally, as $\beta$ goes
below $0.4$, $\xi|_{\eta\tau\approx 6.5}$ rises again while the low
increasing rate is maintained. 

The simulation results imply that the network with the smaller value of $\beta$
tends to take a smaller $\xi$, or be denser in the comoving box. To see
this fact from another aspect, we calculate the effective number of
strings in a virtual box whose volume is $\HH^{-3}$, which hereafter we
refer to as ``horizon-sized box.'' Focusing on a string, it occupies an
area $\xi^2$ on the surface perpendicular to the string by
definition. Hence, considering a cross-sectional surface of the
simulation box, one can imagine that $\HH^{-2}/\xi^2$ strings pass
across the surface. According to this naive expectation, we estimate the
effective number of strings in the horizon-sized box by
%
\begin{equation}
 N_{\rm eff} = \frac{1}{\HH^2\xi^2}.
 \label{eq:Neff}
\end{equation}
%
If $\xi$ behaves approximately as $\xi(\tau) = c\tau + \xi_0$ where $c$
and $\xi_0$ are constants and we neglect the oscillatory behavior, 
$N_{\rm eff}$ at the late time becomes
%
\begin{equation}
 N_{\rm late} = \lim_{\tau\to\infty}N_{\rm eff} =  \frac{1}{c^2},
 \label{eq:Nlate}
\end{equation}
%
where we used $\HH = 1/\tau$ in the radiation-dominant Universe. 
This equation indicates that the network density in the horizon-sized
box is determined by the increasing rate of $\xi$, and also
the correspondence to the fact that the number of strings in the
horizon-sized box is constant in the scaling regime. 
We estimate the ensemble average,
$\langle c\rangle$, using the data obtained by all realizations, and then we
obtain the late-time number of strings in the simulation box, $N_{\rm late}$. 

The details of the procedure to obtain $N_{\rm late}$
are explained in Appendix~\ref{appsec:smoothing}.
Briefly speaking, we first remove the oscillatory behavior of $\xi$ in
each realization, and then using the least-square method, we fit the data
in the range of $\tau_{\rm start}\leq\tau\leq\tau_{\rm end}$ to the 
ansatz $\xi = c\tau+\xi_0$ where $\tau_{\rm start}\eta=6.5$ and
$\tau_{\rm end}$ is dependent on the value of $\beta$. How to choose
$\tau_{\rm end}$ is discussed in Appendix~\ref{appsec:check}.

Figure~\ref{fig:Nlate} shows $N_{\rm late}$ against $\beta$. 
The error bars are due to the variation of $c$ in each realization.
We found that $N_{\rm late}$ becomes obviously larger for 
$\beta < 0.8$, while it is almost constant and only a few 
strings exist for $\beta \geq 0.8$. Although $\beta=0.4$
realized the maximum number of strings, it might be premature to
conclude that the network with $\beta=0.4$ becomes densest in the horizon-sized box.
As discussed in Appendix \ref{appsec:check}, the reliable range of
simulation data for $\beta<0.4$ is relatively short.
In order to extend the reliable range, many more spatial resolutions are
required to resolve the steep change of the gauge field around
the string core, indicated by $m_A\to\infty$ for $\beta\to 0$, as discussed
in Appendix \ref{appsec:vortex}. 

\begin{figure}[!ht]
\centering{
 \includegraphics[width=8cm]{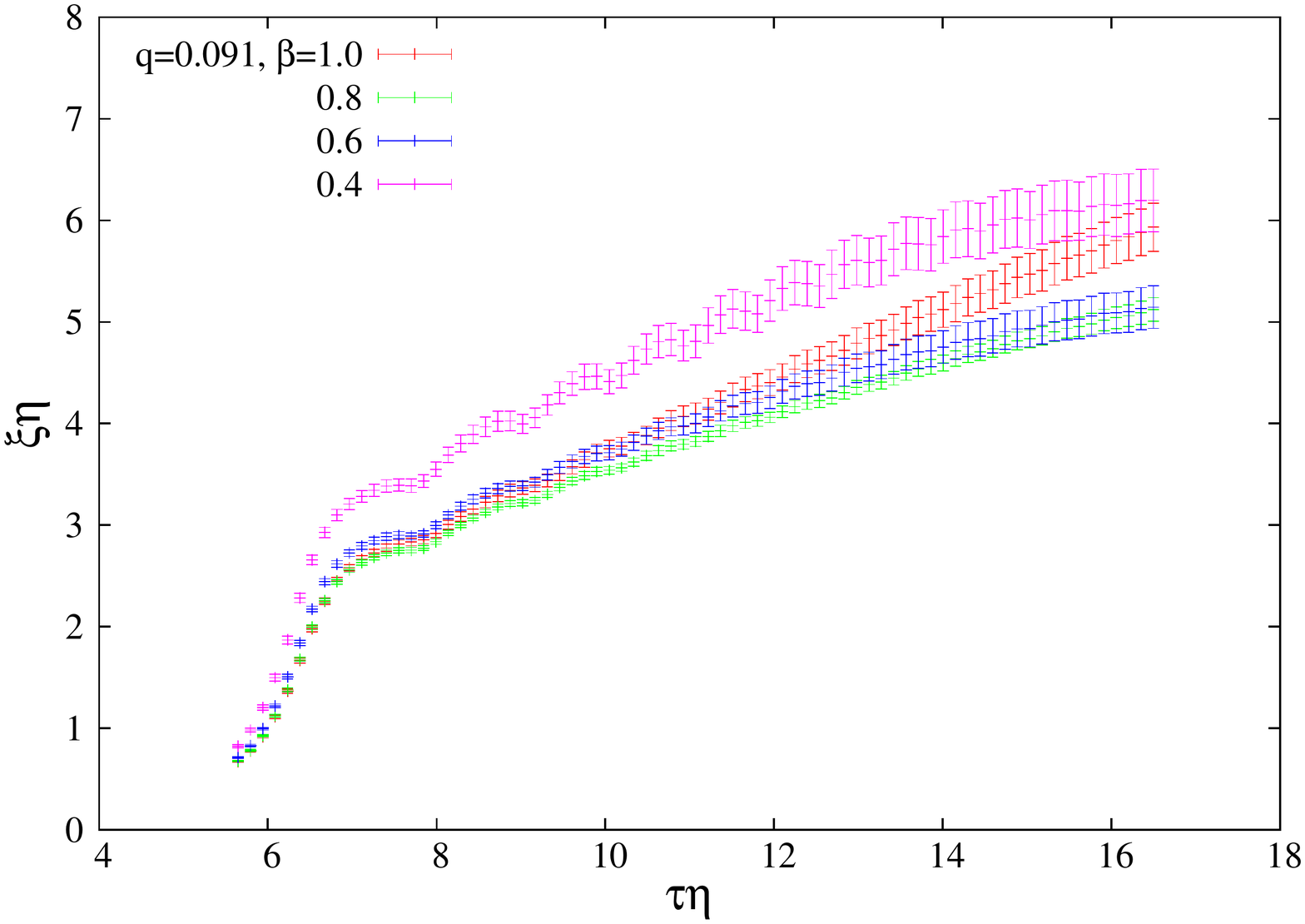}
 \includegraphics[width=8cm]{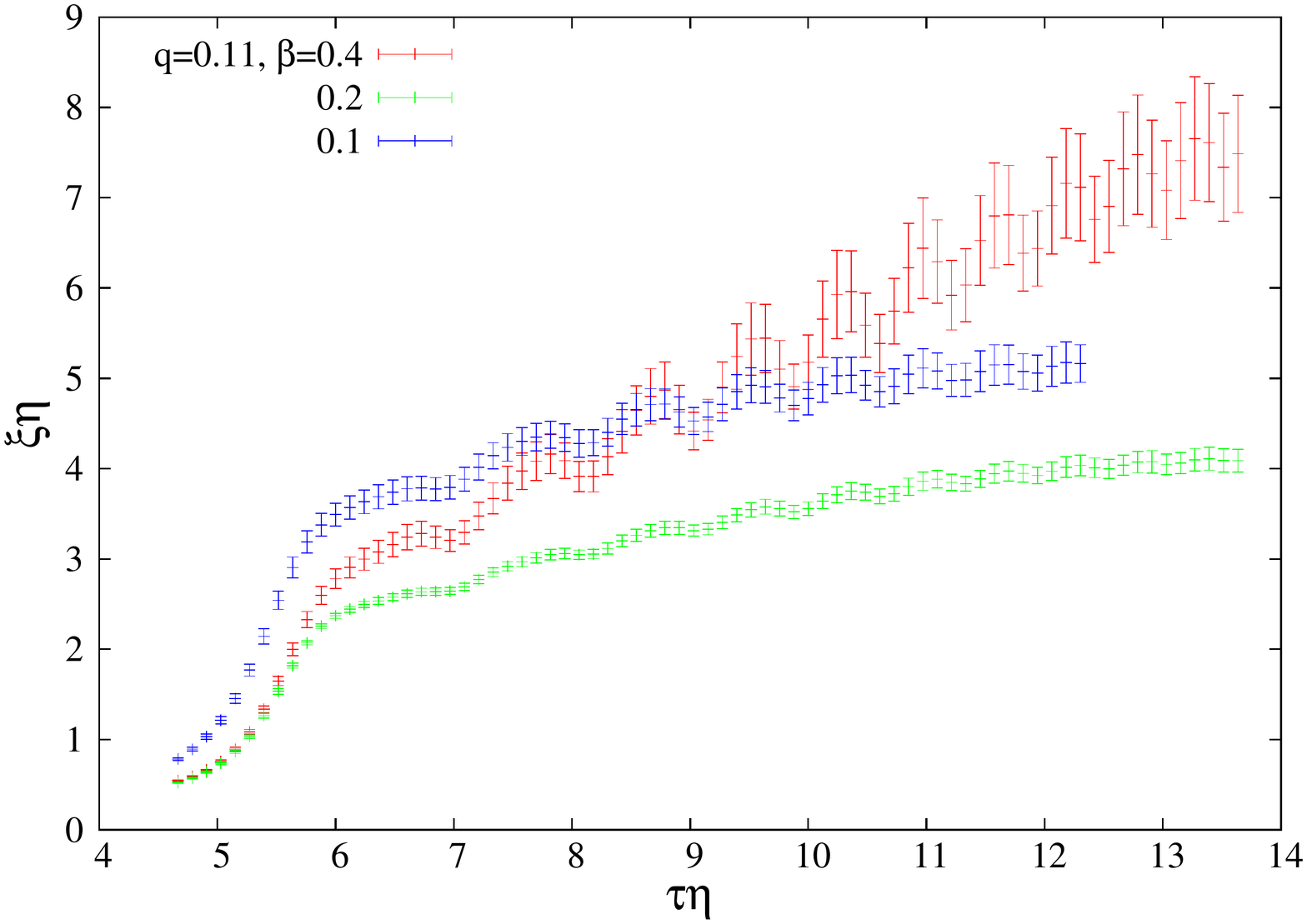}
}
\caption{The correlation length with $q=0.091$ (left) and 
$q=0.11$ (right). A lower energy scale of the phase transition (left)
 results in a larger critical value of $\beta$ realizing the smallest
 $\xi$. In the present case, $q=0.091$ provides the critical $\beta$
 between $0.6$ and $1.0$, whereas a higher phase transition with
 $q=0.11$ provides a smaller critical $\beta\sim 0.2$.} 
\label{fig:diffH}
\end{figure}

Back to the network density in the comoving box, we performed additional
simulations for $q=0.11$ and $0.091$ with fixed $\gamma$ where $q$ is
defined in Sec.~\ref{sec:setup}, and check whether
the critical value of $\beta$ realizing the densest network in the
comoving box depends on $q$. For the large value of
$q$, the phase transition takes place at the earlier
Universe since the transition temperature is given by 
$T_c=\sqrt 3\eta=\sqrt{3}q\gamma\Mpl$. In Fig.~\ref{fig:diffH}, 
the phase transition at higher energies with $q=0.11$ provides
a smaller critical $\beta$ around $0.2$, whereas the critical $\beta$ is
not so clear for $q=0.091$, but would be around $0.8$.
These facts imply
that the string network density in the comoving box is strongly related
to the dragging effect by the cosmic expansion. 

\begin{figure}[!ht]
\centering{
 \includegraphics[width=10cm]{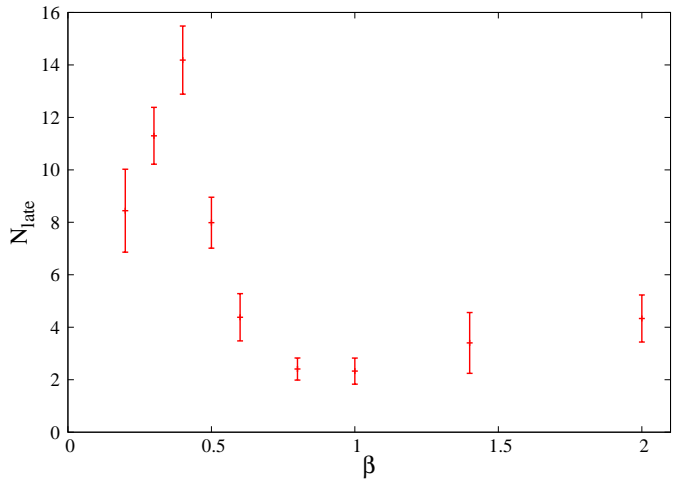}
}
\caption{The effective number of strings in a horizon-sized box
 at the late time defined in Eq.~(\ref{eq:Nlate}). The error bars are
 originated from the variation of $c$ and the increasing rate of $\xi$ for each realization.
 }
\label{fig:Nlate}
\end{figure}

\subsection{PRS algorithm}

\begin{figure}[!ht]
\centering{
 \includegraphics[width=10cm]{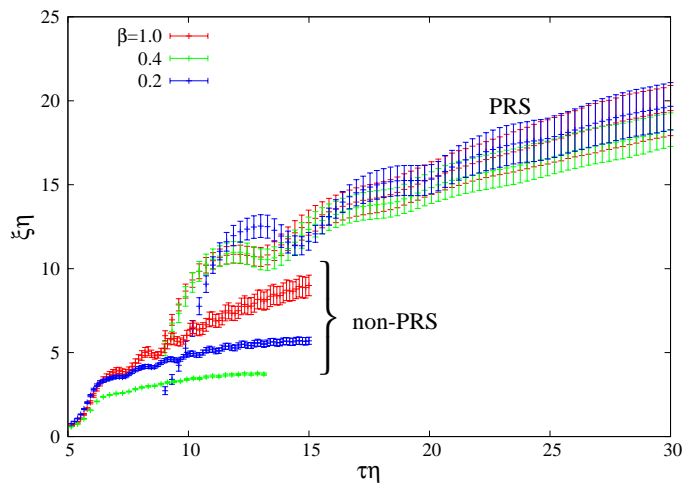}
}
\caption{The correlation length of simulated string network with and
 without the PRS  algorithm.}
\label{fig:PRS}
\end{figure}

Finally, we investigate the effect of the PRS algorithm, given in
Eq.~(\ref{eq:PRS}), on the correlation lengths of the simulated Type-I
string network. We choose $\beta=0.2$ and $0.4$, and also
$\beta=1.0$ as a reference. Their correlation lengths
are shown in Fig.~\ref{fig:PRS}. 
In this figure, we also plotted the corresponding results shown in
Fig.~\ref{fig:corr}. As we mentioned before, there is a
strong dependence of $\xi$ on the value of $\beta$. In contrast, if we
use the PRS algorithm, the $\beta$ dependence completely disappears, and
thus it seems that we can obtain the universal value of the gradient of
$\xi$ and $N_{\rm late}$.

Focusing on the case with the critical coupling (red line), 
the gradient of $\xi$ is almost insensitive to whether we use the PRS or
not.
This feature has been expected since there would be no interactions, or
sufficiently weak, between strings in this case, except at the impact
points of the reconnection process. Therefore, even if we vary the value
of the coupling constants $\lambda$ and $e$ in time, the characteristics
of the whole network are not so affected.

In contrast, this is not the case with the Type-I strings, which have
the intrinsic string-string interactions. To use the PRS algorithm
for them corresponds to weakening the interactions on purpose as time 
proceeds. Hence, the network would lack its characteristics depending on the value of
$\beta$. Consequently, we would like to claim the need to carefully
apply the PRS algorithm to the cases, except those with the critical coupling.
Note that, for those with the PRS, the scaling regime starts later 
than those without the PRS. This is because $\lambda$ is decreased at
and after the phase transition. Smaller $\lambda$ means that the
potential becomes more flat and thus it takes more time for the field to
sufficiently relax and then to satisfy
$V(\varphi;T)<V(0;T)$. 

\section{Conclusion and discussion}
\label{sec:conclusion}

We numerically studied the formation and time evolution of Type-I
cosmic-string networks in the Abelian-Higgs model by three-dimensional
lattice simulations in a box with $N=1024^3$ grid. Figure~\ref{fig:energy} was useful to
check that the network actually enters the scaling regime. Then we
particularly focused on the dependence of the correlation length $\xi$
on the parameter $\beta$. In the Type-I regime ($\beta<1$), 
the gauge field plays an important role for the interaction between strings.
As seen in Fig.~\ref{fig:corr}, we found that the time
dependence of the correlation length is strongly dependent on the value
of $\beta$. More concretely, we found that there seems to be a critical
value of $\beta$ with which the string network becomes densest in the
expanding Universe,\footnote{Note that the critical value of $\beta$
defined here does not mean the critical coupling, namely, $\beta=1$.} and that this critical
value becomes smaller, if the energy scale of the phase transition
becomes higher; see Fig.~\ref{fig:diffH}. Furthermore, we found that the
effective number of strings in a box with the volume $\HH^{-3}$, as
defined in Eq.~\eqref{eq:Neff}, is almost constant for
$\beta\gtrsim 0.8$, and the number tends to suddenly increase 
for $\beta\lesssim 0.8$; see Fig.~\ref{fig:Nlate}.
The figure also indicates that the number of
strings has a peak at $\beta=0.4$. However, it would be premature to
conclude that $\beta=0.4$ actually realizes the densest network
since the reliable range of simulation data is not sufficiently wide 
due to the shortage of the spatial resolution of our simulations for
$\beta<0.4$. 

The critical value of $\beta$ seems to depend on the energy scale of the
phase transition. We found that the phase transition at higher energies
provides a smaller critical $\beta$, whereas the value becomes larger
if the phase transition takes place at lower energies. This fact implies
that the critical $\beta$ realizing the densest network is determined
not only from the strength of the gauge interaction, but also from the
environmental effect, namely, the cosmic expansion. 
In order to clarify the origin of the critical $\beta$, 
it would be needed to deeply investigate the string-string interaction
in the Friedmann background.

So far, field-theoretic simulations of string network formation have been
performed with the Press-Ryden-Spergel (PRS) algorithm where $\lambda$
and $e$ are varied in time to maintain the constancy of the comoving
width of a string. This algorithm is effective for
the lattice simulations in the expanding Universe since the expanding
lattice spacing can forever, in principle, follow the width of the
strings. In order to investigate the validity of this algorithm for the
Type-I strings, we have also performed the simulations with the PRS
algorithm. As a result, the interesting properties mentioned above
completely disappeared, and hence we cannot find any differences among the
results with different values of $\beta$. This result indicates that the
PRS algorithm should not be applied to Type-I strings, if one focuses on the
epoch soon after the phase transition where the string-string
interaction would be strong.

However, there is a subtlety in the connection between the two results
with and without the PRS algorithm in Fig.~\ref{fig:PRS}. Naively
thinking, we can speculate that, at the sufficiently late time, the mean separation of the
strings would become large enough for them to terminate the interactions
with each other. This fact would mean that the correlation length
evolves along with the results with the PRS algorithm shown in
Fig.~\ref{fig:PRS} at the late time, since the change of string width
must be negligible at the sufficiently late time.
In other words, it is expected that the gradient of $\xi$ without the
PRS would become larger at some time when the string-string interaction
can be neglected, and then the gradient becomes similar to that with the
PRS. Unfortunately, with our present computer resources, we could not
follow the simulations up to such a transition point, and thus this is
still an open question.

\begin{acknowledgments}
We thank Professor~A.~Vilenkin for giving useful comments on this work. T.~H.
is supported by JSPS Grant-in-Aid for Young Scientists (B) No.~23740186,
and also by MEXT HPCI Strategic Program. Y.~S. is supported in part by
MEXT through Grant-in-Aid for Scientific Research on Innovative Areas
No.~24111701. K.~T. is supported by JSPS Grant-in-Aid for 
Young Scientists (B) No.~23740179, by MEXT through Grant-in-Aid for
 Scientific Research on Innovative Areas 
No.~24111710, and partially by JSPS Grant-in-Aid for Scientific Research (B)
No.~24340048.
D.~Y. was supported by Grant-in-Aid for JSPS Fellows No.~259800.
\end{acknowledgments}

\appendix

\section{Convergence check of numerical results}
\label{appsec:check}

\begin{figure}[!ht]
\centering{
 \includegraphics[width=5.5cm]{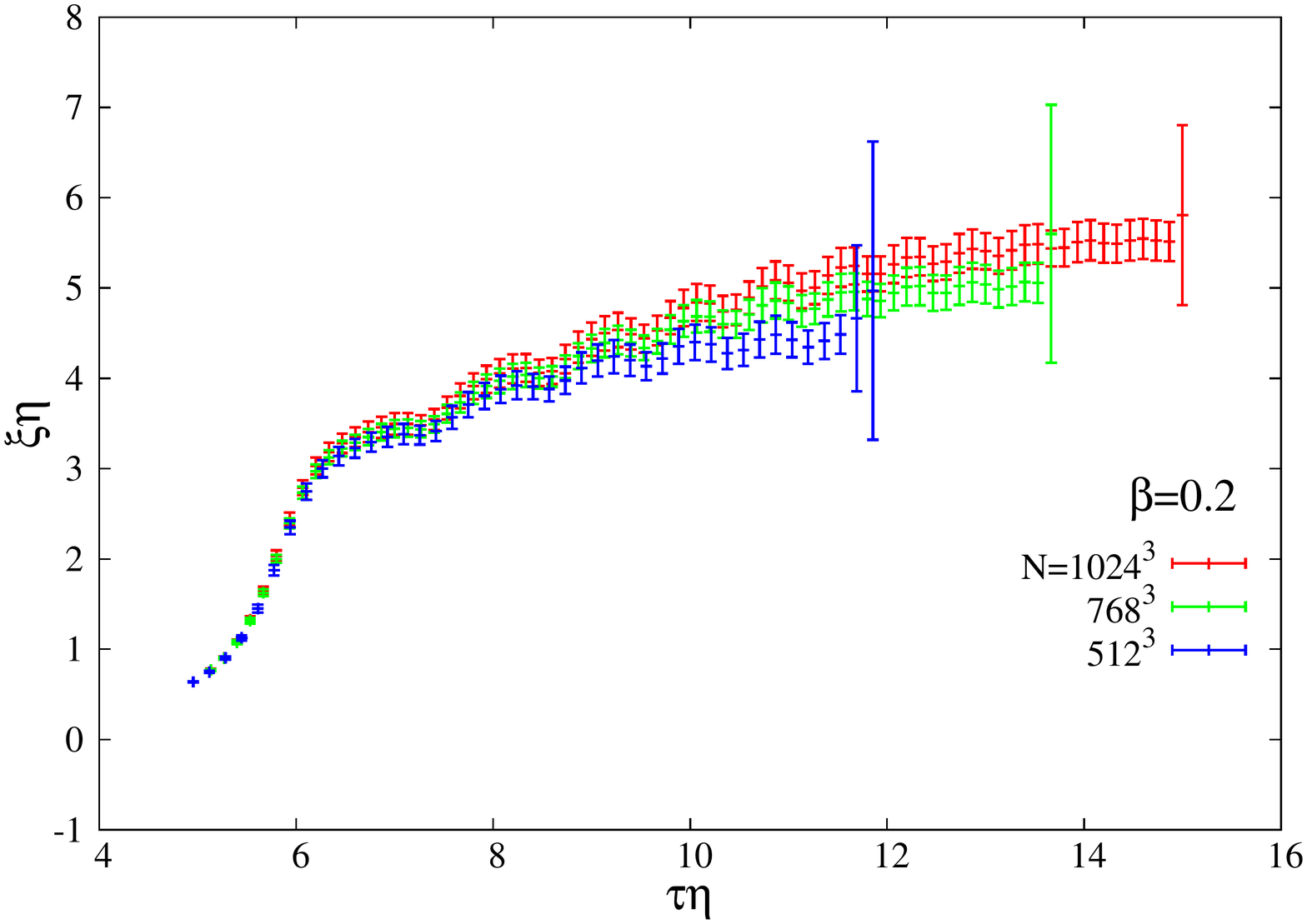}
 \includegraphics[width=5.5cm]{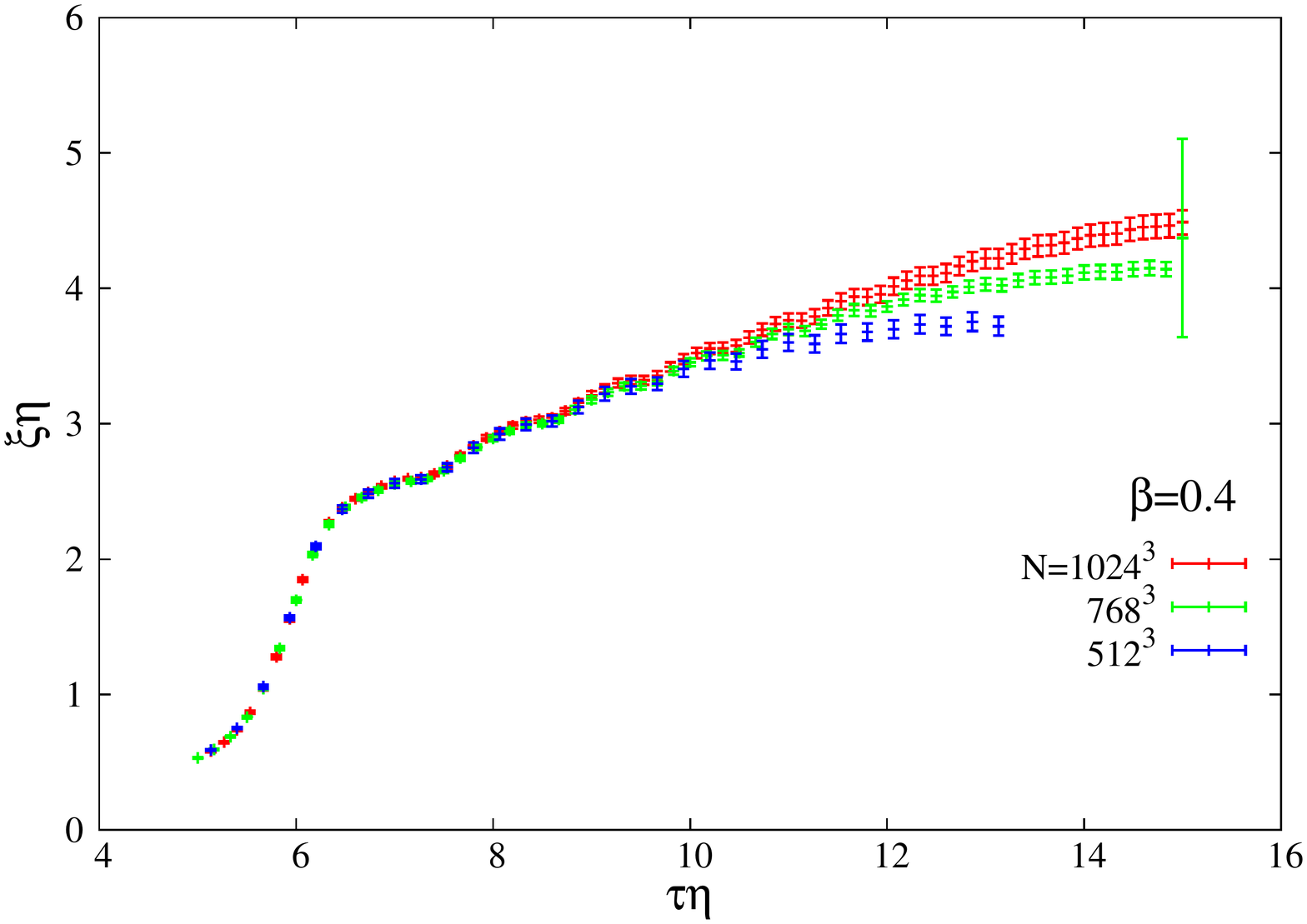}
 \includegraphics[width=5.5cm]{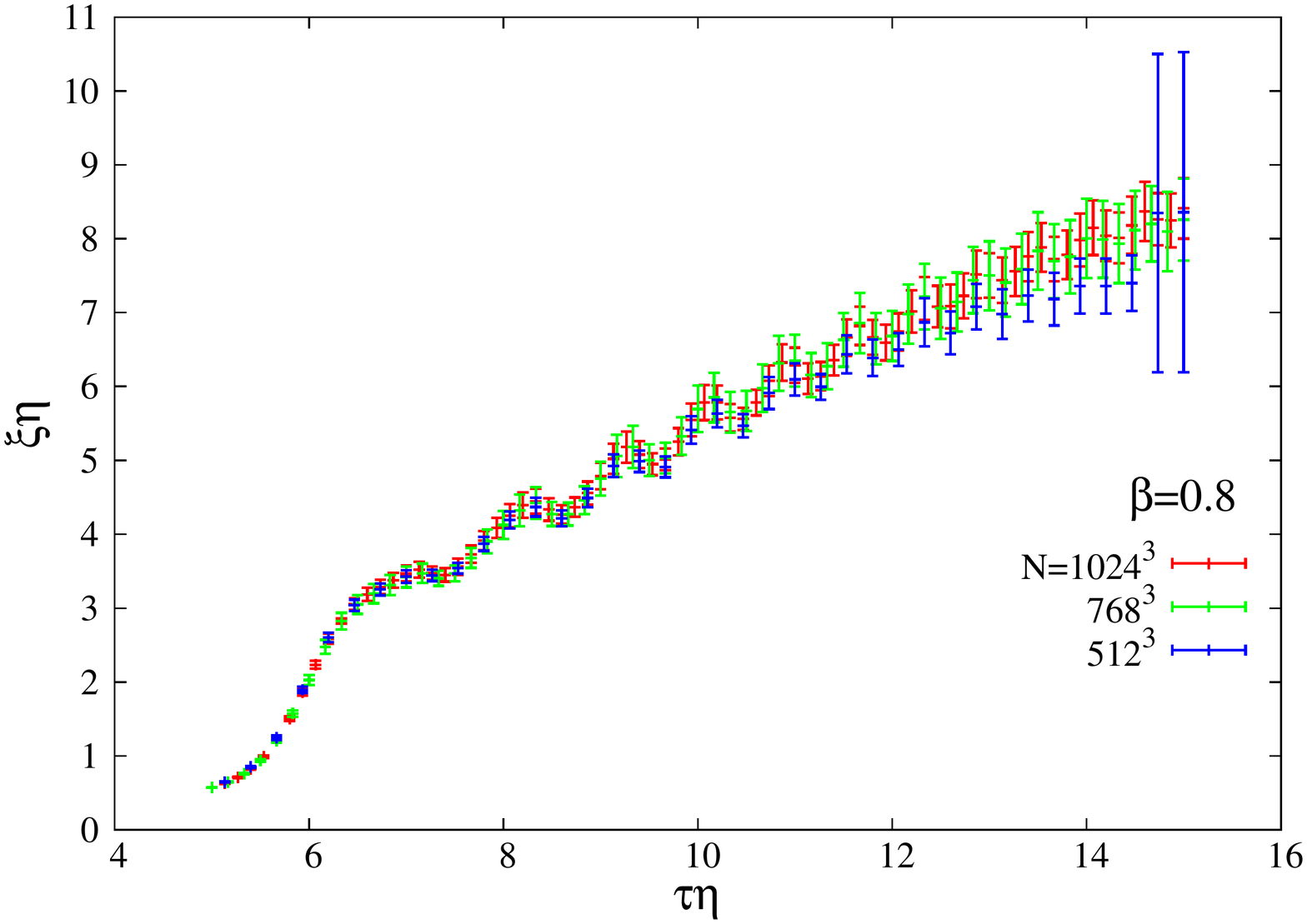}
}
\caption{The resolution dependence of the estimated correlation length
 with $\beta=0.2$ (left), $\beta=0.4$ (center) and
 $\beta=0.8$ (right). We fixed the box size, $L$, and varied the grid size as
 $N=1024^3$ (red), $768^3$ (green) and $512^3$ (blue).}
\label{fig:resolution}
\end{figure}

We check the robustness of our numerical results to the
resolution. Figure~\ref{fig:resolution} shows that the correlation
length with $\beta=0.2$, 0.4 and 0.8 when we vary the number of grids with a fixed
box size, $L$. In the comoving coordinate, a string seems to become thinner
in time, and thus the simulation is broken down when the grid can no
longer resolve the string. This fact reflects that the end time of each
simulation becomes later as $N$ is increased. Moreover, just before the
breakdown, $\xi$ tends to be flat, while it grew almost linearly. Therefore,
the reliable results would be obtained only in the region where two
results with different resolutions overlap. 

Due to this shortage of resolutions at the late time, we use only the
relatively reliable part of simulation data in the finite time range,
$\tau_{\rm start}\leq \tau \leq \tau_{\rm end}$, when we estimate the
gradient of $\xi$ in Sec.~\ref{subsec:correlation} 
or Appendix~\ref{appsec:smoothing}. For all cases, we fix 
$\tau_{\rm start}=6.5\eta^{-1}$ corresponding to the starting time of
the scaling regime, and basically $\tau_{\rm end}=15\eta^{-1}$ which is
the end time of simulations. From the results in Fig.~\ref{fig:resolution}, we choose 
$\tau_{\rm end}=13\eta^{-1}$ for $\beta=0.4$, and 
$\tau_{\rm end}=12\eta^{-1}$ for $\beta=0.2$. 

\section{Static vortex width}
\label{appsec:vortex}

\begin{figure}[!ht]
\centering{
 \includegraphics[width=10cm]{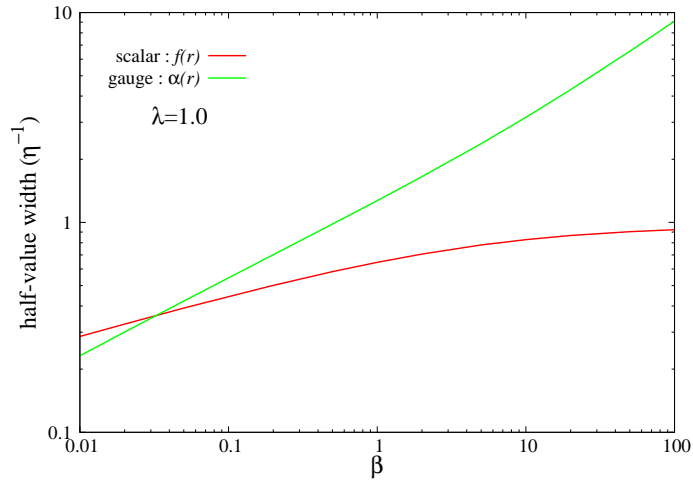}
}
\caption{The $\beta$ dependence of the half-value width of functions
 $f(r)$ and $\alpha(r)$ for $n=1$ strings with $\lambda=1.0$.} 
\label{fig:vortexwidth}
\end{figure}

Consider the axial-symmetric string configuration in the Minkowski
spacetime, so-called Abrikosov-Nielsen-Olesen vortex \cite{Abrikosov:1956sx,Nielsen:1973cs}.
According to Ref.~\cite{Vilenkin}, using the cylindrical
coordinates $(r,\theta,z)$ that originated from the center of the string, the
scalar field and the gauge field can be represented by the following
forms:
%
\begin{align}
 \varphi(\xx) &= \eta e^{in\theta}f(r), \\
 A_\theta(\xx) &= \frac{n}{e}\alpha(r),\qquad A_r=A_z=0,
\end{align}
%
where $n$ is the winding number of the string.
Substituting them into Eqs.~(\ref{eq:eqfullphi}) and (\ref{eq:eqfullA})
with $g_{\mu\nu}={\rm diag}(-1,1,r^2,1)$, and
neglecting the time dependence,
the governing equations for $f(r)$  and $\alpha(r)$ are given by
%
\begin{gather}
 \frac{d^2f}{d\tr^2} + \frac{1}{\tr}\frac{df}{d\tr} -
 \frac{n^2f}{\tr^2}(\alpha-1)^2 - \lambda f(f^2-1) = 0, \\
 \frac{d^2\alpha}{d\tr^2} - \frac{1}{\tr}\frac{d\alpha}{d\tr} 
  - 2e^2f^2(\alpha-1) = 0,
\end{gather}
%
where $\tr=r\eta$ and we neglect the temperature-dependent terms in the
potential $V(\varphi)$. The boundary conditions are given as
$f(r),\alpha(r)=0$ for $r=0$ and  $f(r),\alpha(r)\to 1$ for 
$r\to \infty$. With these conditions, we solved the above equations
numerically. Then we calculated the half-value widths of $f(r)$ and
$\alpha(r)$, the value of $r$ satisfying $f(r)=1/2$ or $\alpha(r)=1/2$,
for the various $\beta$ as the estimator of the string
width. Figure~\ref{fig:vortexwidth} shows the $\beta$ dependence of the
half-value widths for $n=1$ strings with $\lambda=1.0$. Clearly the
string cores consisting of the scalar field and the gauge field get
thin for Type-I strings ($\beta<1$). In other words, the large $e$ with
a fixed value of $\lambda$ or the small $\lambda$ with a fixed value of
$e$ produces thin strings. In particular, the half-value width of the
gauge field is more strongly dependent on $\beta$ than that of the
scalar field. This property requires the finer resolution of the
computational domain, particularly for $\beta<1$.

%

\section{Estimation of gradient of $\xi$, and $N_{\rm late}$}
\label{appsec:smoothing}

The following is the flow chart to estimate $N_{\rm late}$ 
defined in Eq.~(\ref{eq:Nlate}) from the raw
simulation data of $\xi$ shown in Fig.~\ref{fig:corr}.
\begin{itemize}
\item Determining the reliable range of $\xi$ data, $\tau_{\rm end}$,
      discussed in Appendix~\ref{appsec:check}.
\item Smoothing $\xi$ to remove the oscillatory behavior.
\item Fitting each smoothed $\xi$ to a linear function of $\tau$ to
      obtain the gradient of $\xi$, and
      averaging all realizations to obtain the expectation value of the gradient
      and its variance.
\end{itemize}

First of all, according to the convergence check against the spatial resolution
discussed in Appendix~\ref{appsec:check}, we determined the reliable
range of $\xi(\tau)$ data, $\tau_{\rm end}$. Next, let us consider the smoothing process
for the raw data of $\xi(\tau)$ containing the oscillations. Our final
goal is to fit $\xi$ to a linear function such as 
$\xi(\tau) = c\tau+\xi_0$. Hence the smoothing process should not 
affect the gradient $c$. The simplest treatment would be the averaging
with neighbor points in the time domain. Defining $\xi_i \equiv \xi(\tau_i)$ for
$i=1\ldots N$ and $\tau_1=\tau_{\rm start}, \tau_N=\tau_{\rm end}$, 
this averaging can be described as
%
\begin{equation}
  \xi_1 = {\rm fixed}, \qquad
  \xi_i^{(j+1)} = \frac{\xi_{i-1}^{(j)}+\xi_{i+1}^{(j)}}{2}, \qquad
  \xi_N = {\rm fixed}, 
  \label{eq:smoothing}
\end{equation}
%
where $\xi_i^{(j)}$ represents the $j$ th smoothed data, and 
$\xi_i^{(0)}$ is the original raw data. We repeat
this process until $j \leq j_{\rm max}$. This formula can be derived by
approximating the equation $d^2\xi/d\tau^2|_{\tau=\tau_i}=0$ with the
second-order central difference formula. 

\begin{figure}[!ht]
\centering{
  \includegraphics[width=10cm]{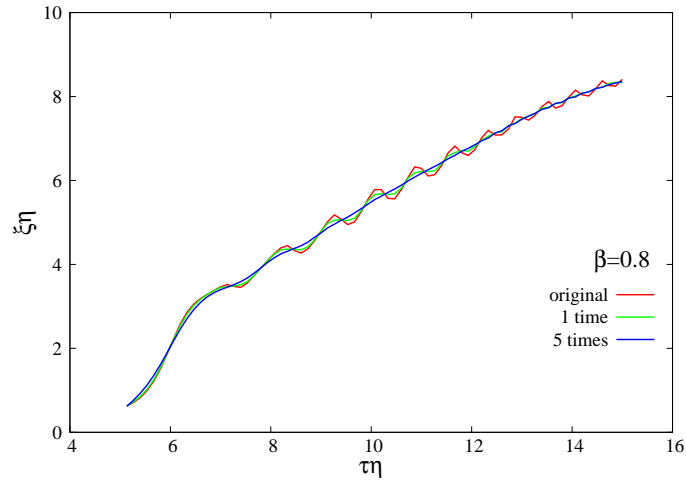}
}
\caption{The red line is the original data of $\xi(\tau)$ for
 $\beta=0.8$. The green and blue lines are the resultant smoothed data
 with one time iteration and five times iterations, respectively.}
\label{fig:smoothing}
\end{figure}

In Fig.~\ref{fig:smoothing}, the blue line represents the resultant
smoothed $\xi$ with $j=5$ times iterations, while the red line
represents the original data. It is clear that only the
oscillatory behaviors are successfully removed, and the global gradient
does not change during this process.

Finally, we fit the linear regime of the smoothed $\xi$ 
in $\tau_{\rm start}\leq \tau\leq \tau_{\rm end}$
to a linear function, $\xi(\tau)=c\tau+\xi_0$.
According to this procedure, we obtain ten independent values of $c$ 
from the ten sets of simulation data
for a given $\beta$. Then we can calculate the
expectation value of $c$, $\langle c\rangle$, and its variance, $\sigma_c^2$.
Finally, using the definition of $N_{\rm late}$ given in
Eq.~(\ref{eq:Nlate}), we plot Fig.~\ref{fig:Nlate},
where the error bar indicates $\sigma_\xi = |\partial N_{\rm
late}/\partial c|\sigma_c=2\sigma_c/c^3$.

\end{document}